\documentclass[12pt]{article}
\usepackage{graphicx, subfigure}
\usepackage{amssymb}
\usepackage{amscd}
\usepackage{amsmath}
\usepackage{appendix}
\usepackage{verbatim}
\usepackage{epstopdf}

\begin{document}
\begin{titlepage}

{\hbox to\hsize{\hfill August 2019 }}

\bigskip \vspace{3\baselineskip}

\begin{center}
{\bf \Large

Supersymmetric Naturalness Beyond MSSM}

\bigskip

\bigskip

{\bf  Archil Kobakhidze and Matthew Talia$^1$ \\ }

\smallskip

\begin{center}
	{ \textit{\small{}$^{1}$ ARC Centre of Excellence for Particle Physics
			at the Terascale, }\\
		\textit{\small{} School of Physics, The University of Sydney, NSW
			2006, Australia }\\
		\textit{\small{} E-mails: archil.kobakhidze,
			matthew.talia@sydney.edu.au }\\
		\textit{\small{}}}
	\par\end{center}{\small \par}

\bigskip

\bigskip

\bigskip

{\large \bf Abstract}
\end{center}
\noindent 
The experiments at the Large Hadron Collider (LHC) have pushed the limits on masses of supersymmetric particles beyond the $\sim$TeV scale. This compromises naturalness of the simplest supersymmetric extension of the Standard Model, the minimal supersymmetric Standard Model (MSSM). In this paper we advocate that perhaps the current experimental data are actually hinting towards the physics beyond MSSM. 
To illustrate this, we treat the MSSM as a low-energy limit of a more fundamental yet unspecified theory at a scale $\Lambda$, and compute the fine-tuning measure $\Delta$ for generic boundary conditions with soft SUSY breaking parameters and various cut-off scales. As a general trend we observe reduction in fine-tuning together with lowering $\Lambda$. In particular, perfectly natural [$\Delta \lesssim \mathcal{O}(10)$] theories with a multi-TeV spectrum of supersymmetric particles that are consistent with all current observations can be obtained for $\Lambda \sim  \mathcal{O}(100)$TeV. The lowering of the fine-tuning for large cut-off scales can also be observed in theories exhibiting special quasi-fixed point behaviour for certain parameters. Our observations call for a more thorough exploration of possible alternative ultraviolet completions of MSSM.
   
\end{titlepage}

\section{Introduction}

Supersymmetry (SUSY) is a very attractive theoretical framework for physics beyond the Standard Model as it represents a unique non-trivial extension of relativistic invariance and provides a unified description of particles with different spin-statistics. As an important by-product, supersymmetric quantum field theories exhibit improved short-distance scale behaviour. Namely, the notorious quadratic divergences are completely absent even in theories with softly broken SUSY. This feature motivates studies of the SUSY extension of the Standard Model (MSSM) with additional supersymmetric particles in the mass range $\sim 100-1000$ GeV, since such a theory would naturally incorporate the electroweak scale without the need to fine tune quantum corrected parameters. 

Contrary to these theoretical expectations, there has not been any evidence of SUSY particles at the LHC. With some simplified assumptions, the experiments exclude gluinos and first and second generation squarks with masses up to $\sim 2$ TeV \cite{gluino-atlas}. Taking this at face value, this compromises naturalness of the electroweak scale - a fine-tuning of $\Delta \gtrsim 300$ is required to accommodate the correct electroweak scale.

It should be clear however, that the above conclusion in no way falsifies SUSY as a theoretical framework, but rather the MSSM, in its particular realization (see also the recent related discussion in \cite{Ross:2017kjc}). In fact, we would like to advocate here that the failure of the natural MSSM may indicate physics beyond the MSSM. There are two major ways that the parameters involved in fine-tuning can be modified due to the new physics. First is through the modification of the renormalisation group (RG) running and second is through the modification of boundary conditions, e.g., due to the enhanced symmetries in the fundamental theory which are traced down to the low-energy theory. In this paper we take the attitude that the MSSM is a low-energy approximation of a more fundamental yet unspecified theory at a scale $\Lambda$. We than compute the standard Barbieri-Giudice \cite{Barbieri1988} fine-tuning measure $\Delta$ for various scales $\Lambda$, assuming arbitrary boundary conditions on MSSM parameters defined at $\Lambda$. This way we parametrize our ignorance of the fundamental theory as well as the effects of higher-dimensional irrelevant operators which are expected to be important at energies $\Lambda$. As a general trend we observe the reduction of fine-tuning with lowering $\Lambda$ from $\Delta \sim {\cal O}(100)$ to $\Delta \sim {\cal O}(10)$, which may hint towards physics beyond the MSSM at a scale $\sim 100$ TeV\footnote{In Ref. \cite{Cassel:2009ps}, the reduction of fine-tuning was also observed within the effective MSSM with high dimension operators in the Higgs sector.}. This is roughly consistent with the observations made previously in \cite{casas2015}. We also discuss an example of quasi-fixed point running of parameters, which results in a low fine-tuning for large cut-off scales. These examples motivate further search of specific extensions of the MSSM which result in natural electroweak scale. 

The paper is organized as follows. In Section \ref{chap1}, we describe the measure of fine-tuning in the context of the MSSM and subsequently the relationship to the parameter RGEs. In Section \ref{chap2} we perform a general scan over the MSSM, computing the fine-tuning measure along with relevant experimental constraints. Similarly in Section \ref{chap3} we scan over a narrower region of parameter space corresponding to an MSSM quasi-infrared fixed point resulting in low fine-tuning. In Section \ref{concl} we present our conclusions.

\section{Supersymmetric naturalness \label{chap1}} 

We consider the MSSM as an effective low-energy approximation of an unspecified ultraviolet theory with a cut-off scale $\Lambda$. The relevant and marginal operators in the effective theory are those of the MSSM superpotential,   
\begin{equation}
W=\bar{u}{\bf y_u}QH_u + \bar{d}{\bf y_d}QH_d + \bar{e}{\bf y_e}LH_d + \mu H_u H_d~,
\end{equation}
where ${\bf y_{u,d,e}}$ are the 3x3 Yukawa matrices in flavor space, the supersymmetric $SU(3)\times SU(2)\times U(1)$ gauge interactions, and the standard MSSM soft SUSY breaking terms. The effective running parameters evaluated at the cut-off scale $\Lambda$ we identify as the "fundamental" parameters of the effective MSSM. 

Minimization of the tree-level potential gives the following relation between the Z-boson mass, $m_Z$, and the low energy soft breaking masses $m_{H_u, H_d}$ and the supersymmetric $\mu$ parameter:
\begin{equation}
\frac{m^2_Z}{2}=\frac{m^2_{H_d}-m^2_{H_u}\tan^2\beta}{\tan^2\beta-1}-\mu^2 \simeq -m^2_{H_u}-\mu^2~.
\label{1}
\end{equation}
Assuming $|m_{H_d}| \lesssim |m_{H_u}|$, the last approximate equation in (\ref{1}) holds to a very good accuracy for $\tan\beta \gtrsim 3$.  Hence, $m_{H_u}$ and $\mu$ at low-energies must be adjusted in a way to reproduce the Z-pole mass $m_Z\simeq 91$ GeV. This adjustment is natural if not very sensitive to the variation of "fundamental" parameters at $\Lambda$. The quantitative measure of this sensitivity is the Barbieri-Giudice fine-tuning parameter \cite{Barbieri1988}:
\begin{equation}
\Delta = \max \left\{\left| \frac{a_i}{m^2_Z}\frac{\partial m^2_Z}{\partial a_i} \right| \right \}~, \label{FTmeasure}
\end{equation}
where the $a_i$ run over the "fundamental" parameters of the effective low-energy MSSM.

The relation between low energy and "fundamental" parameters are defined by the solution of the respective RG equations and matching conditions at the cut-off scale $\Lambda$. The latter can only be computed if the ultraviolet completion of the MSSM is known. Since we are working within the effective field theory framework, we parameterize our ignorance of the ultraviolet physics by considering an unconventional and arbitrary (within certain range) values of the 20 "fundamental" parameters of the MSSM, which are potentially the most relevant for computing the low-energy parameters for different values of $\Lambda$.   

The running of the supersymmetric parameter $\mu$ exhibits a fixed-point at $\mu=0$ and therefore if taken small ($\mu \sim m_Z$) at $\Lambda$, it will stay small at low-energies. Small $\mu$ is therefore natural. However, for pure scalar mass parameters, such as $m_{H_u}$, such a behaviour is atypical due to the additive contribution of heavy particle masses to the corresponding beta-function. More specifically, 
\begin{equation}
\frac{d}{dt}m^2_{H_u}=\frac{1}{16\pi^2}\left[3|y_t|^2X_t-6g^2_2|M_2|^2-\frac{6}{5}g^2_1|M_1|^2+\frac{3}{5}g^2_1S\right]~, 
\label{RG1}
\end{equation}
where $t=\ln\left(\Lambda^2/\tilde\mu^2\right)$ (here $\tilde \mu$ an arbitrary renormalization scale) and  
\begin{eqnarray}
&X_t&=2(m^2_{H_u}+m^2_{Q_3}+m^2_{\bar{u}_3}+|A_t|^2)~, \\
&S&\equiv m^2_{H_u}-m^2_{H_d}+Tr[\bf{m^2_Q-m^2_L-2m^2_{\bar{u}}+m^2_{\bar{d}}+m^2_{\bar{e}}}]~.\nonumber
\end{eqnarray}
As a result, the low-energy parameter $m^2_{H_u}$ is sensitive to variations of different mass parameters, and if the sparticle spectrum is heavy, the fine-tining measure (\ref{FTmeasure}) is generically large.

One can think of two ways to reduce the required fine-tuning in models with large sparticle masses. First, one assumes that the physics beyond the MSSM enters at a low enough scale $\Lambda$ such that the "fundamental" parameters do not evolve significantly when running down to low energies. In this case, if the fundamental theory is such that no significant fine tuning is required to satisfy the minimization condition (\ref{1}), the RG running cannot destabilize the relation (\ref{1}). We confirm this by numerical analysis - the required fine tuning is significantly reduced for low $\Lambda$, even for a rather heavy spectrum of sparticles. 

Alternatively, if one assumes that $m^2_{H_u}$ dominates over other mass parameters at high energies, the RG equation (\ref{RG1}) takes the approximate form:    
\begin{equation}
\frac{d}{dt}m^2_{H_u}= \frac{6y_t^2}{16\pi^2} m_{H_u}^2~,
\label{RG2}
\end{equation}
 which (similar to $\mu$ parameter) exhibits an infrared fixed-point at $m_{H_u}^2 =0$. This observation motivates us to scan a specific region of "fundamental" parameters in section \ref{chap3}. In accord with the expectation, we observe significant reduction in fine-tuning measure for a large $\Lambda$ and heavy sparticles.

From the observation in (\ref{RG1}), we can determine another infrared fixed-point by defining the sum:
\begin{equation}
\Sigma=m^2_{H_u}+m^2_{Q_3}+m^2_{\bar{u}_3}+|A_t|^2~,
\end{equation}
from which we can compute the beta function for $\Sigma$ in the limit that all other mass parameters are subdominant
\begin{equation}
\frac{d}{dt}\Sigma=\frac{3 y^2_t}{4\pi^2} \Sigma - \frac{2}{\pi^2}g_3^2 M^2_3~.
\end{equation}
For $M_3 \rightarrow 0$ this expresses an infrared fixed-point at $\Sigma = 0$. However, since $g_3$ and $M_3$ increase in the infrared, once can expect a significant positive contribution to $\Sigma$. We confirm in the numerical analysis the correlation between the gluino mass $M_{\tilde{g}}$ and fine tuning.

Before we proceed with our numerical analysis, we note that the infrared quasi-fixed point solution in the MSSM in which the top-Yukawa coupling $y_t$ is kept large at the grand unified scale, are well known \cite{Hill1981,Ross1981}. More specifically, upon computing the beta-functions for $y_t$ and $g_3$ up to one-loop and without electroweak contribution, one finds an infrared stable point at $y^2_t/g^2_3=7/18$ of the corresponding RG equations. This procedure has also been carried out for other couplings and soft-masses in the MSSM \cite{Allanach1997,Yeghi1999,Jurcisin1999}. Here we allow more generic variation of fundamental parameters at the high energy scale $\Lambda$ rather than focusing on model-dependent correlations (such as in grand unified theories) among them.  

\section{Parameter Scan} \label{chap2}

In this section we present our results for a generic scan of parameters and different values of $\Lambda$. We retain a full 20 parameter version of the MSSM and perform a broad random scan over the following space:
\begin{eqnarray}
-3000\,\text{GeV} <& M_1,M_2 &< 3000\,\text{GeV} \nonumber \\
&M_3 &< 2000\,\text{GeV} \nonumber \\
-(3000)^2\,\text{GeV}^2 <& m^2_{H_u},m^2_{H_d} &< (3000)^2\,\text{GeV}^2 \nonumber \\
& m^2_{i_{1,2}} &< (3000)^2\,\text{GeV}^2 \nonumber \\
& m^2_{i_3} &< (3000)^2\,\text{GeV}^2 \nonumber \\
-3000\,\text{GeV} <& A_t,A_b,A_{\tau} &< 3000\,\text{GeV} \nonumber \\
1 <& \tan \beta &< 50 \nonumber \\
&sign(\mu)&=\pm 1~.
\end{eqnarray}
where $i=(Q,\bar{u},\bar{d},L,\bar{e})$. The first and second generation scalar soft masses are taken to be degenerate and we assume no flavour mixing at the input scale.

We choose the input scale $\Lambda$ in which the parameters are defined for the following three cases:
\begin{equation}
\Lambda \in \left[10^5,10^{10},10^{16} \right]\,\text{GeV}~.
\end{equation}

We employ full two-loop RGEs using \texttt{SPHENO-3.3.8} \cite{Spheno2003}, combined with \texttt{SARAH} \cite{Staub2014}, in order to compute the MSSM spectrum and fine-tuning measure. The parameters included in the calculation of the fine-tuning measure in Eq. \ref{FTmeasure} are the gaugino masses $M_1,M_2,M_3$, Higgs soft-breaking masses $M^2_{H_u},M^2_{H_d}$, 3rd generation scalar masses $m^2_{Q_3},m^2_{\bar{u}_3},m^2_{\bar{d}_3},m^2_{L_3},m^2_{\bar{e}_3}$, the trilinear couplings $A_t,A_b,A_{\tau}$, and the terms $\mu$ and $B_{\mu}$, all computed at the corresponding scale $\Lambda$. The top (pole) mass is set to 173 GeV. We also compute the DM relic density $\Omega h^2$ and spin-independent WIMP-nucleon cross-section assuming a neutralino DM candidate using \texttt{micrOmegas-4.3.2} \cite{Belanger2002}.

\begin{figure}[h]
	\centering
	\includegraphics[height=0.21\paperheight]{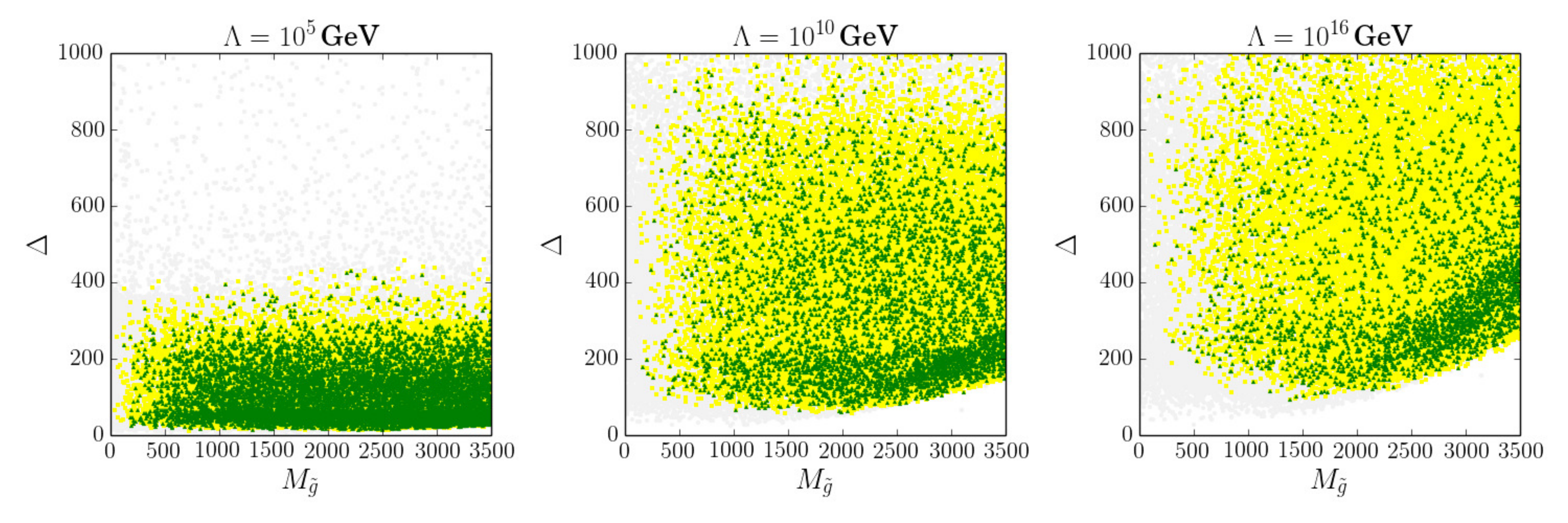}
	\includegraphics[height=0.21\paperheight]{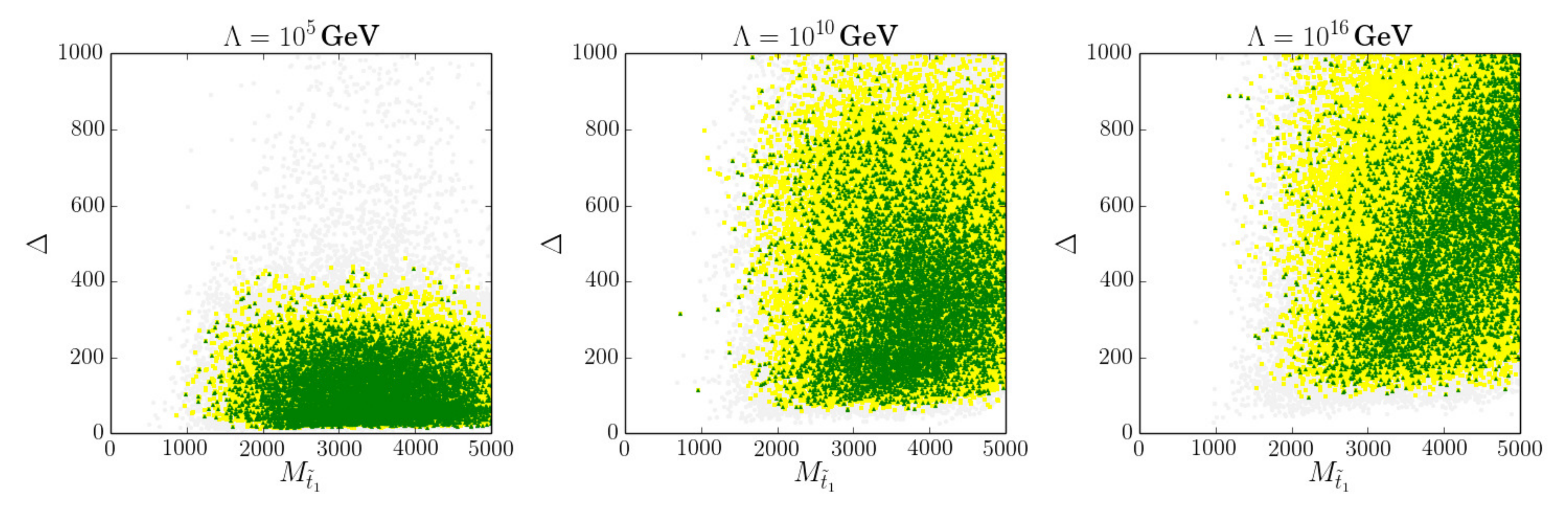}
	\caption{Fine-tuning measure as a function of the gluino mass (top) and lighter stop mass (bottom) for three representative NP scales. Yellow squares contain LEP and Higgs mass constraints as well as $B$-Physics and Higgs precision constraints. The green triangles are a subset of these containing DM relic density and direct detection constraints.}
	\label{general}
\end{figure}

Points which have a vacuum in the electroweak broken phase are chosen which satisfy $\Delta\leq1000$ are subsequently passed through the following constraints:
\begin{itemize}
	\item Direct searches for the slepton and chargino at LEP produce the mass limits on the first two generation sleptons and lightest chargino \cite{pdg}:
	\begin{eqnarray}
	m_{\widetilde{l}_L},m_{\widetilde{l}_R} &>& 100\,\text{GeV} \quad (l=e,\mu)~, \\
	m_{\widetilde{\chi}^{\pm}_1} &>& 105\,\text{GeV}~.
	\end{eqnarray}
	\item We require the lightest Higgs mass in the range $122 < m_h < 127\,\text{GeV}$ \cite{higgs-atlas,higgs-cms},
    \item We require the lightest neutralino $\widetilde{\chi}^0_1$ as the LSP and $m_{\widetilde{\chi}_1^0}>30\,\text{GeV}$ to be consistent with the bound on light MSSM neutralino dark matter \cite{Calibbi:2014lga},
    \item We satisfy the 3 sigma upper-bound on dark matter relic density observed by the PLANCK collaboration \cite{planck} given by $\Omega_{Planck}h^2 = 0.112 \pm 0.006$. For points with underabundant dark matter, we assume there may be some additional contribution from non-thermal candiates, such as the axion.
    \item We use the recent data from XENON1T \cite{Xenon2017} to constrain the points from direct detection experiments, where we rescale the spin-independent cross-section $\sigma^{SI}$ with the observed relic density by $(\Omega h^2/\Omega_{Planck}h^2)$,
    \item We check the bounds from Higgs searches at LEP, Tevatron and LHC implemented using \texttt{HiggsBounds-4.3.1} \cite{Bechtle2015},
    \item We also check important $B$-physics constraints, namely $BR(B\rightarrow X_s\gamma)$ and $BR(B_{S} \rightarrow \mu^{+}\mu^{-})$. The measured values we use are $BR(B\rightarrow X_s\gamma)_{\text{exp}}=(3.55 \pm 0.26)\times10^{-4}$ \cite{HFAG2011} and the upper bound $BR(B_{S} \rightarrow \mu^{+}\mu^{-})_{\text{exp}}<1.08\times10^{-8}$ (95$\%$ CL) \cite{CMSLHCb2011}. These are calculated using \texttt{FlavorKit} \cite{Flav1} as part of the \texttt{SPheno/SARAH} package. Where an upper and lower bound are shown, we constrain our points to within $3\sigma$ of the quoted value.
\end{itemize}
We do not impose constraints from gluino/squark searches from ATLAS and CMS as the limits are model-dependent and would require a dedicated recasting of the collider limits. Besides, there are many cases in which the spectrum may be compressed to easily avoid these LHC search constraints. However, as we will see the LHC constraints on squark and gluino masses can be easily satisfied in many cases.

In Figure \ref{general} we show the dependence on the fine-tuning measure on the gluino mass $M_{\tilde{g}}$ and lighter stop mass $M_{\tilde{t}_1}$. As expected, one finds that when the ``fundamental" parameters are entered at the low scale, chosen at $\Lambda=10^5$ GeV, there is less constraint on a heavier spectrum whilst the electroweak scale still remains natural. Little variation in the parameters from renormalization group evolution even allows for fine-tuning as low as $\Delta_{min} \sim 10$ in this case. Remarkably, this is true for gluinos and stops with multi-TeV masses, well beyond the reach of current experiments.

\begin{figure}[h]
	\begin{center}
		\includegraphics[height=0.25\paperheight]{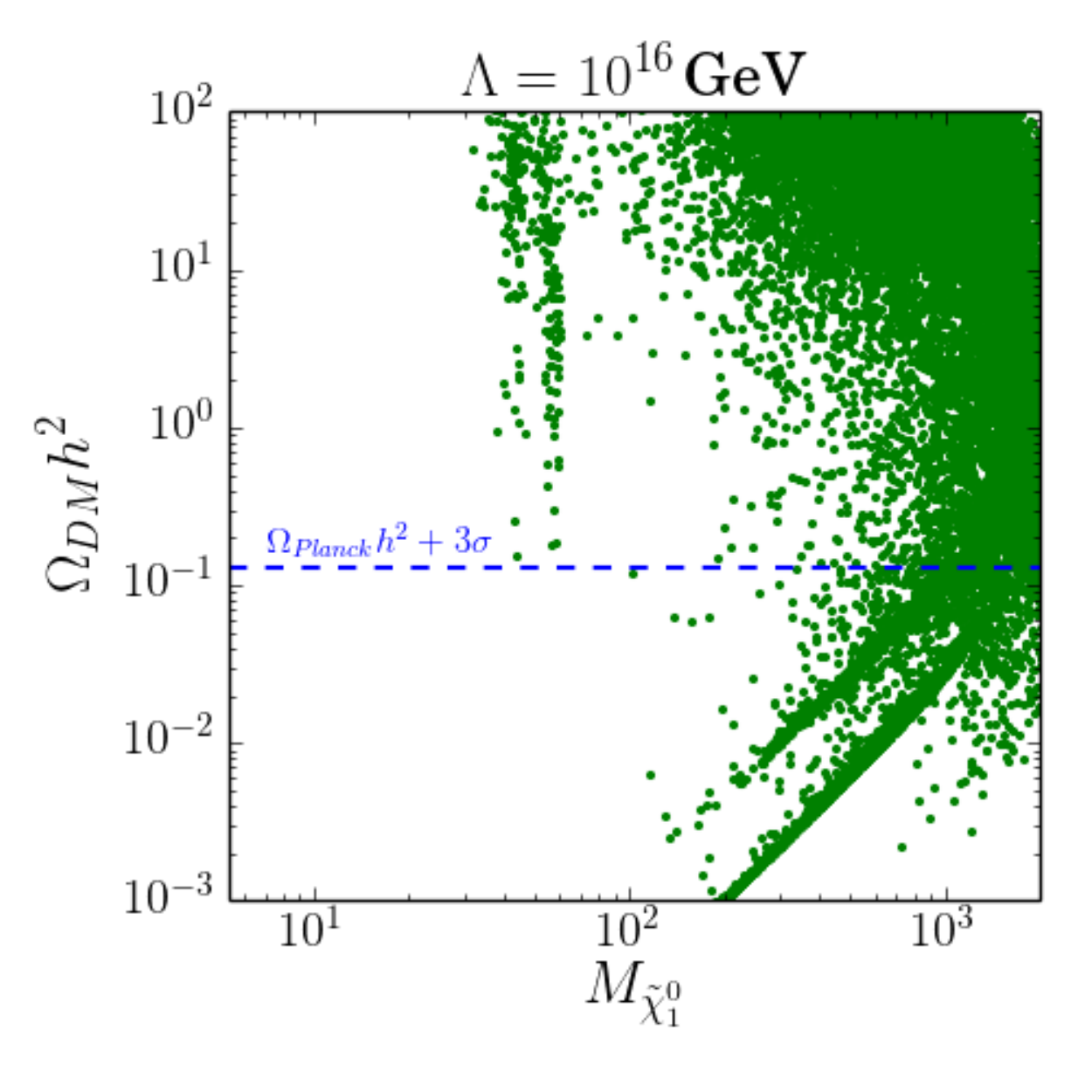}
		\includegraphics[height=0.25\paperheight]{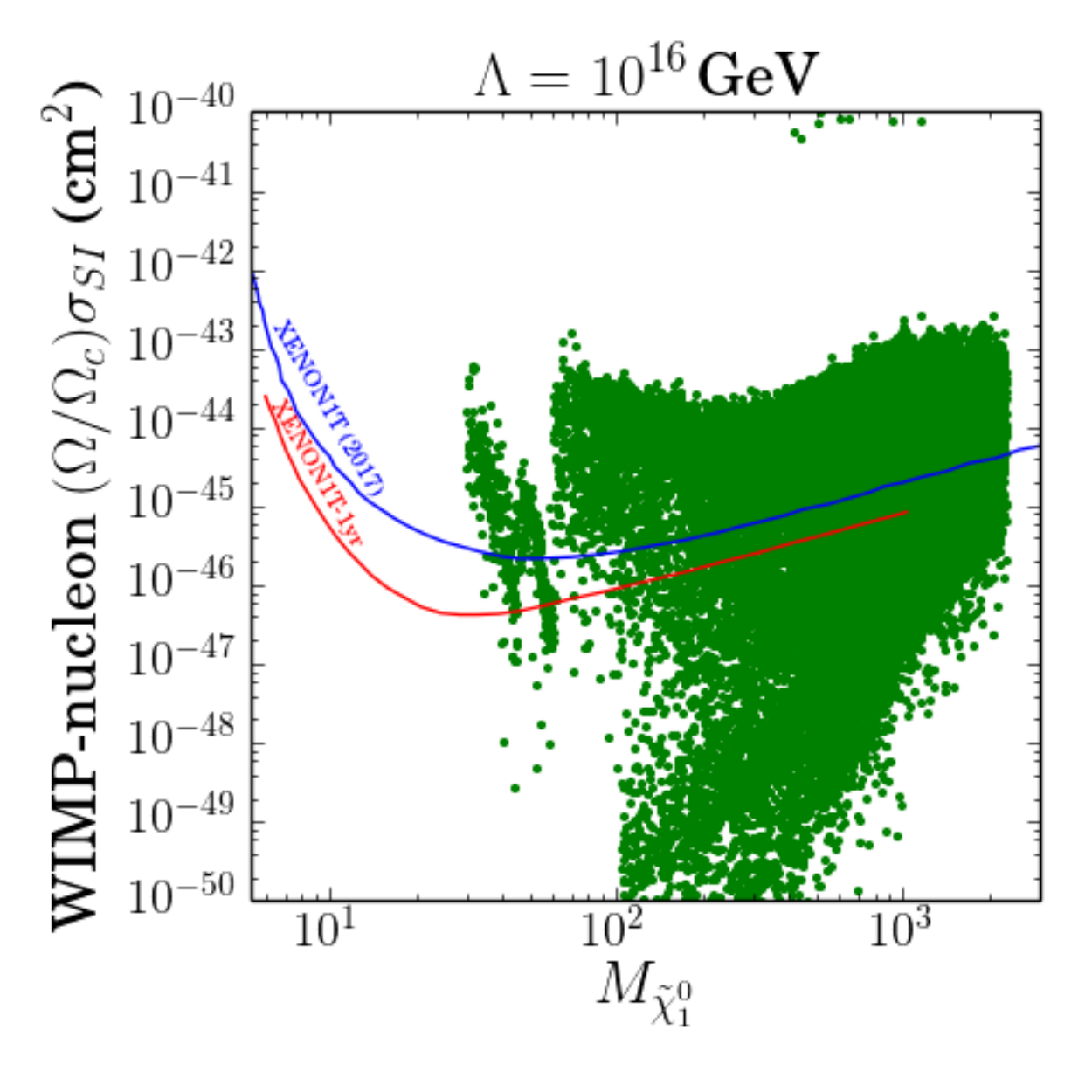}
	\end{center}
	\caption{Left: Relic density $\Omega_{DM}$ as a function of the LSP mass corresponding to the red squares in the rightmost panels of Figure \ref{general}. The dotted line corresponds to the PLANCK measurement of $\Omega_{Planck}h^2 = 0.112 \pm 0.006$ \cite{planck}. Right: WIMP-nucleon spin-independent cross-section as a function of the LSP mass for the points shown in the left panel. Since we allow the LSP to be underabundant after freeze-out, we rescale the cross-section by the factor $\Omega/\Omega_{c}$ where $\Omega_c=\Omega_{Planck}$. The solid lines correspond to the XENON1T 2017 \cite{Xenon2017} and the recent 1 tonne $\times$ year \cite{Xenon2018} results. Similar plots exist for $\Lambda=10^{5}$ and $10^{10}$ GeV.}
	\label{dmplots}
\end{figure}

The constraints on relic density and direct-detection of DM can be satisfied relatively easily, shown in Figure \ref{dmplots}, since the contributions from the electroweakino masses to the RG running of the up-type Higgs mass soft-breaking term is mild. We would also like to stress that the dark matter abundance, besides the microscopic properties, depend crucially on the cosmological evolution of the universe. In particular, the region of parameter space with over-abundant dark matter (ie. a mostly bino-like LSP) shown in Figure \ref{dmplots} can actually be consistent with observation with depopulation mechanisms shown in \cite{Baker2017,Kobakhidze2017} effective in the early universe.

We conclude this section by stressing that perfectly natural theories are possible even for multi-TeV spectra of sparticles, which not only satisfy the current LHC bounds but in some cases are quite beyond the reach of the LHC. This is especially true for the low cut-off at $\Lambda \sim 100$ TeV which motivates further searches for physics beyond the MSSM with a natural electroweak scale.

\section{MSSM Quasi infrared fixed-point and fine-tuning} \label{chap3}

\begin{figure}
	\begin{center}
		\includegraphics[width=0.45\textwidth]{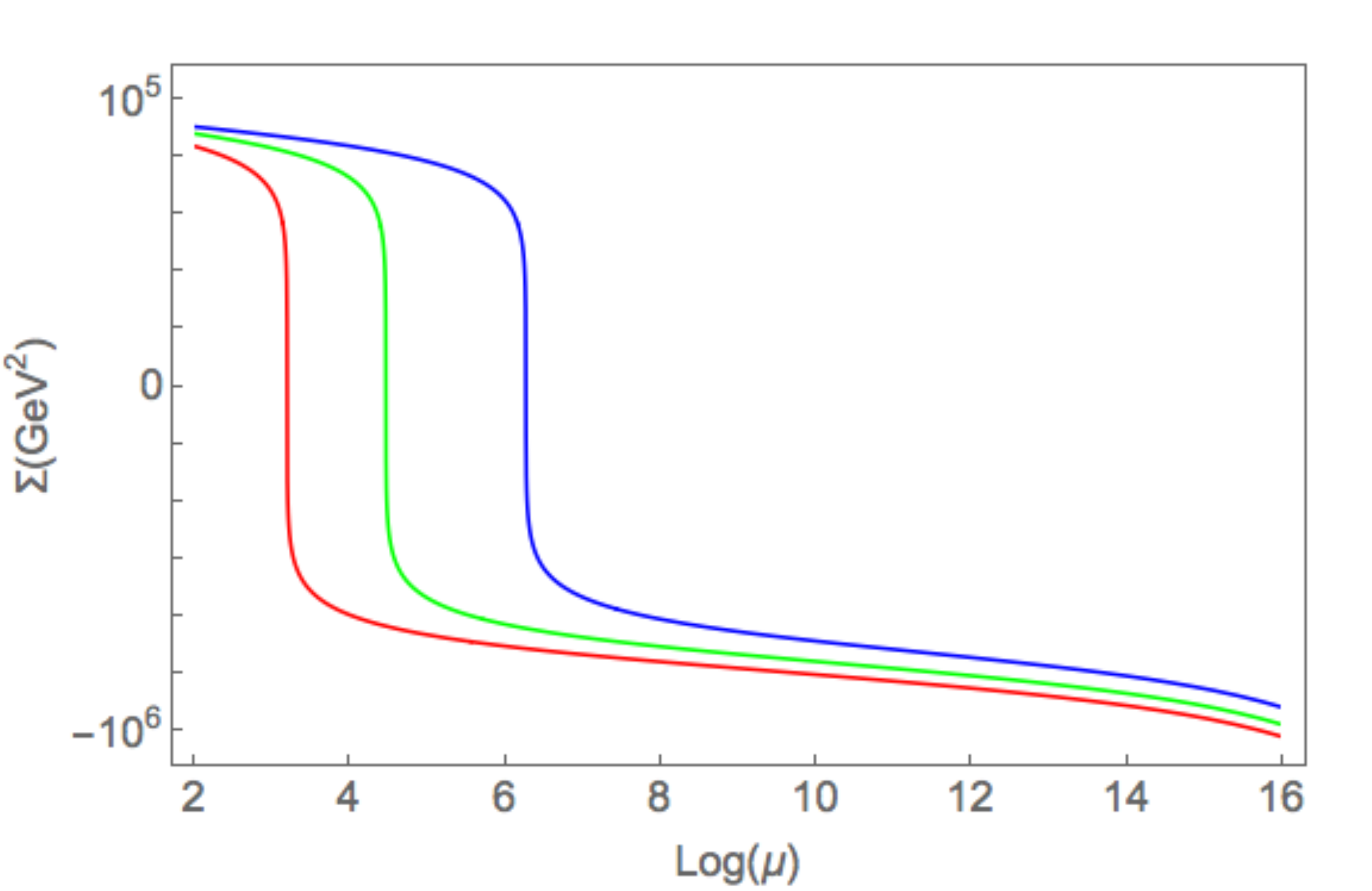}
		\includegraphics[width=0.45\textwidth]{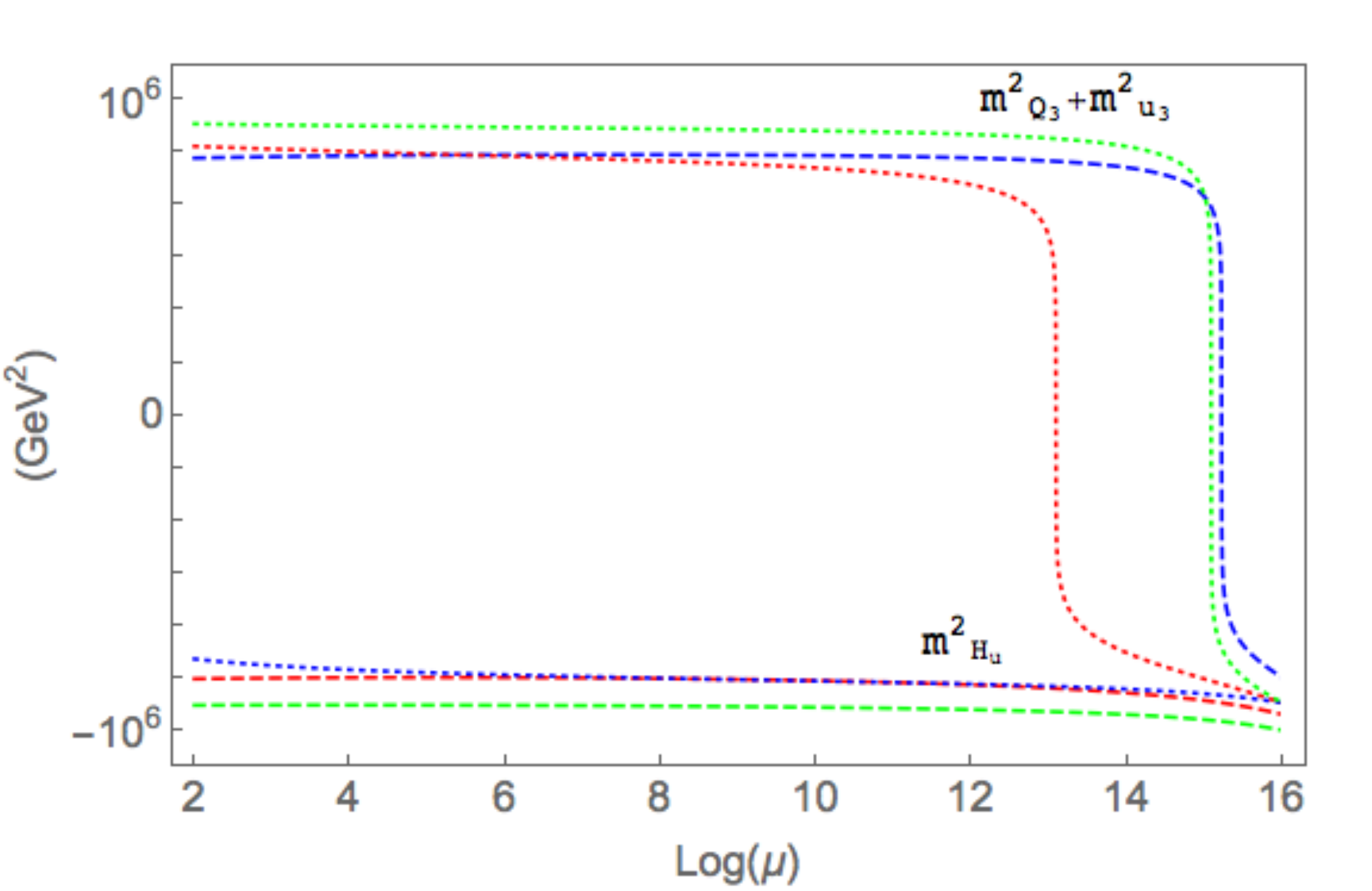}
		\includegraphics[width=0.45\textwidth]{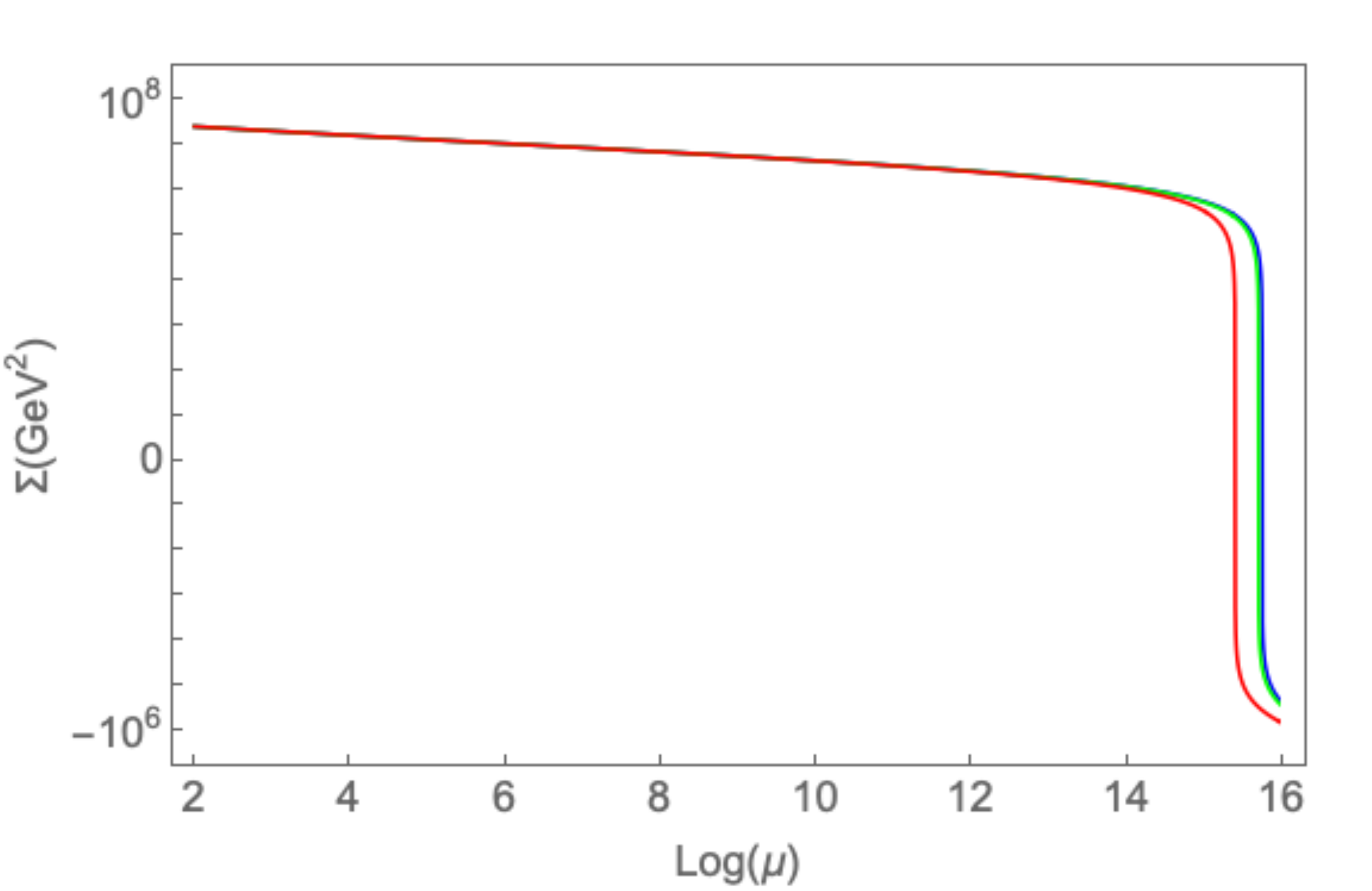}
		\includegraphics[width=0.45\textwidth]{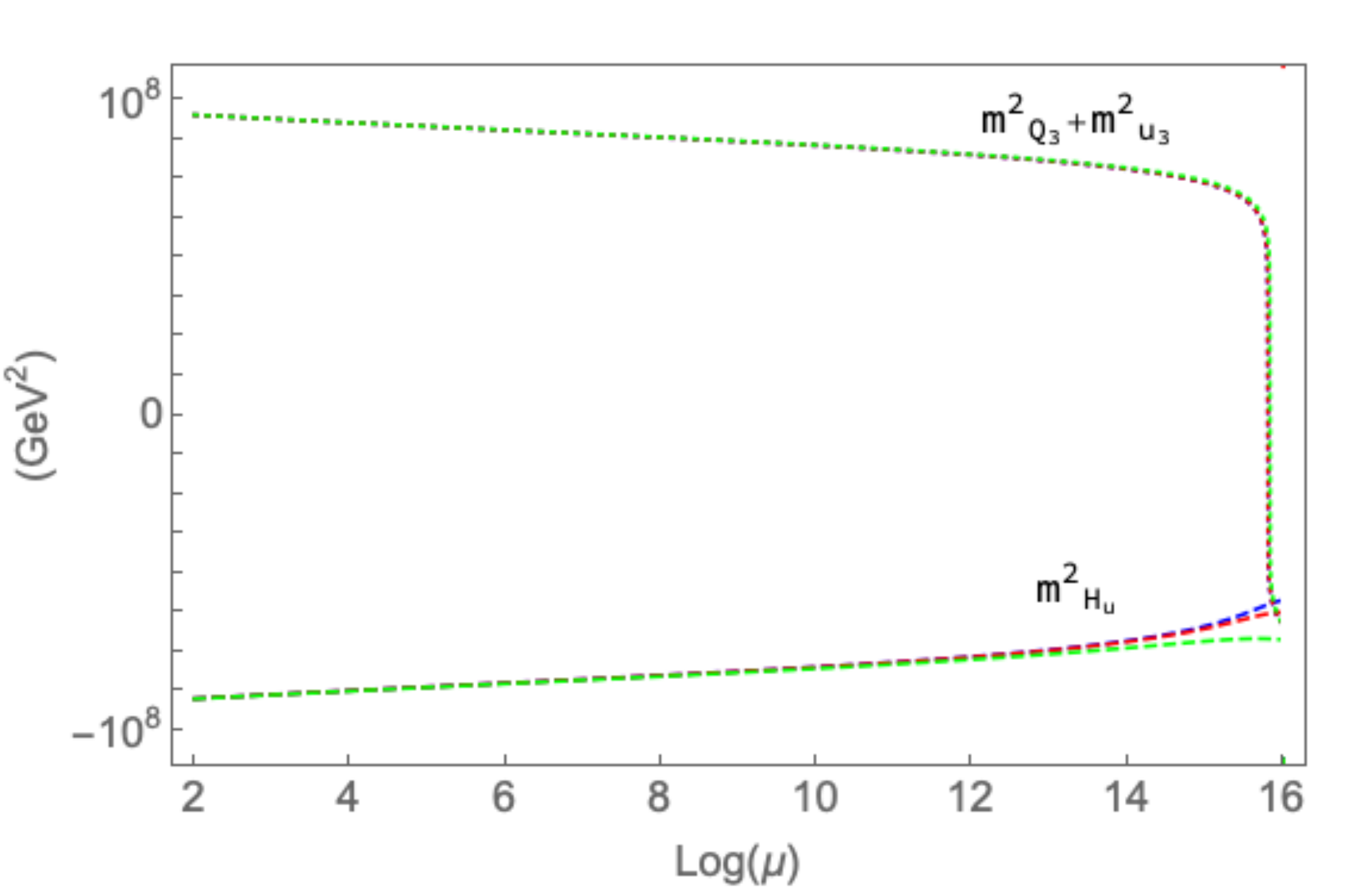}
	\end{center}
	\caption{Evolution of the parameter $\Sigma$ to the infrared fixed-point and soft mass parameters input at $\Lambda=10^{16}$ GeV with $m^2_{Q_3}=m^2_{\bar{u}_3}=-10^5\,\text{GeV}^2$, $A_t=-100$ GeV and $\tan \beta=10$. The three separate curves are shown for different initial values of: Top Row: $M_3=100\,\text{GeV},m^2_{H_u}=-10^5, -5\times 10^5, -10^6\, \text{GeV}^2$. Bottom Row: $M_3=2500\,\text{GeV},m^2_{H_u}=-5\times10^4, -10^5, -5\times10^5\, \text{GeV}^2$. }
	\label{rgeplots}
\end{figure}

In the following, we choose a large top Yukawa coupling, $y_t>1$ at the high-scale to enhance the running of $m^2_{H_u}$. In order to enhance the contribution from $m^2_{H_u}$ at the scale $\Lambda$ we allow it to be dominant over $m^2_i$ where $i$ runs over the scalar mass squared values, excluding the 3rd generation squark soft-masses. The 3rd generation squark masses $m^2_{Q_3}$ and $m^2_{\bar{u}_3}$ tend to de-stabilize $m^2_{H_u}$ as they largely contribute a positive value toward the infrared, leading to a large value of $m^2_{H_u}$ at $M_{SUSY}$. One can avoid this with negative scalar mass-squared parameters at the input scale. Negative stop mass-squared parameters at the GUT scale have been previously studied in some gauge messenger models \cite{Dermisek2006} and the MSSM \cite{Dermi2006}. We present an example of the RGE evolution of these soft mass parameters in Figure \ref{rgeplots}. Most notably, the gluino mass parameter significantly raises the fixed-point value of $\Sigma$ in the infrared, whilst simultaneously enhancing the running of the stop mass parameters into positive values. The fixed-point behavior requires $m^2_{H_u}$ run to more negative values for large $M_3$, increasing the fine-tuning.\footnote{This is analogous to the \textit{"gluino sucks"} effect discussed in \cite{Arvan2014}.}

\begin{figure}
	\centering
	\includegraphics[height=0.2\paperheight]{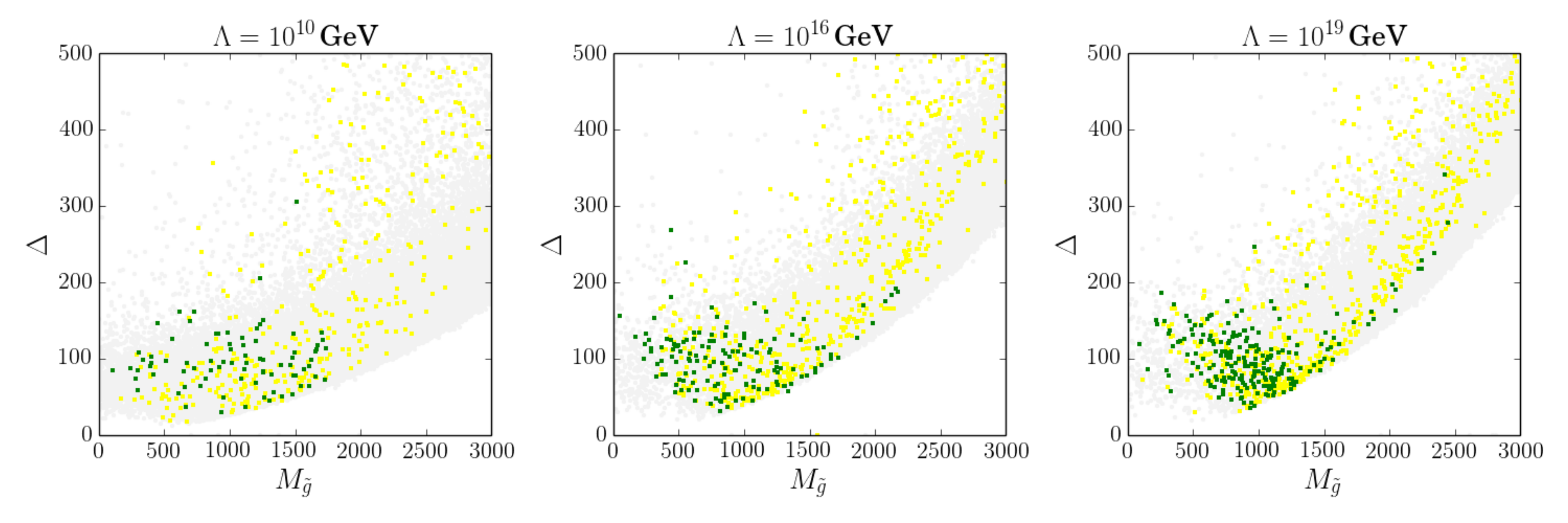}
	\includegraphics[height=0.2\paperheight]{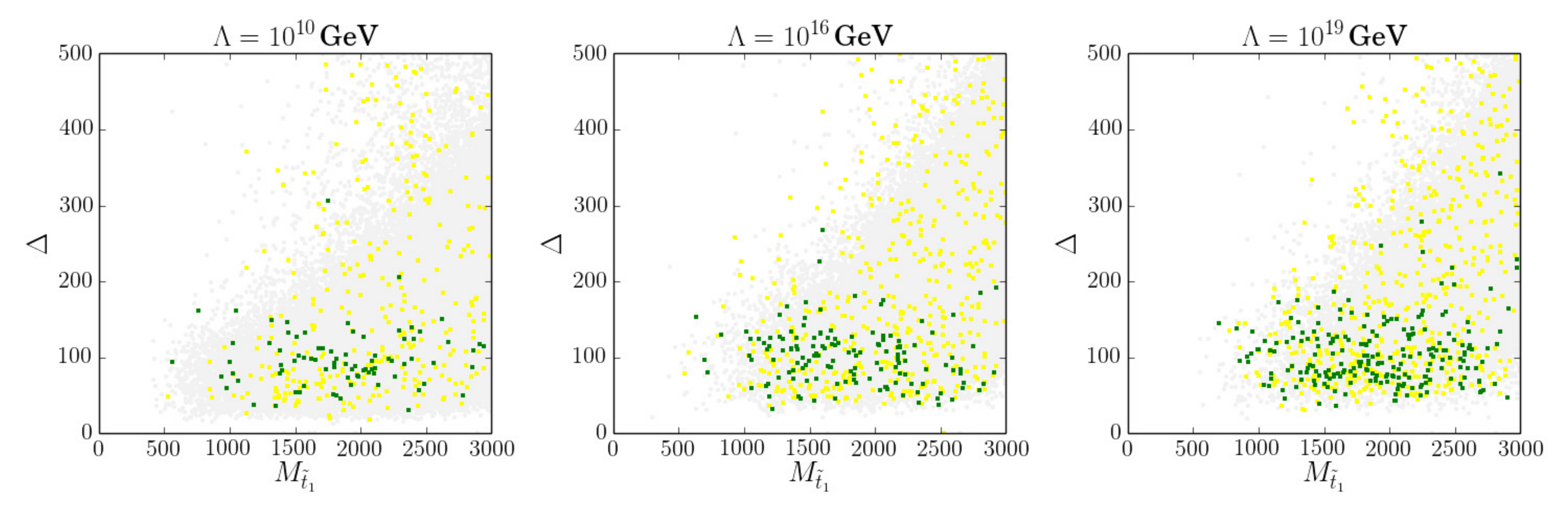}
	\caption{Same as in Figure \ref{general} with higher NP scales and in a narrower scan range supporting the infrared fixed-point behavior for $\Sigma$. Note the reduced range of $\Delta$ values on the y-axis and higher $\Lambda$ values compared to Figure \ref{general}. Small fine tuning of $\Delta < O(100)$ can be achieved even for heavier sparticle masses $> 1$ TeV.}
	\label{QFPplots}
\end{figure}

With this as our motivation, we more precisely scan over the following modified space:
\begin{equation}
\Lambda \in \left[10^{10},10^{16},10^{19} \right]\,\text{GeV}~,
\end{equation}
\begin{eqnarray}
-3000\,\text{GeV} <& M_1,M_2 &< 3000\,\text{GeV} \nonumber \\
&M_3 &< 3000\,\text{GeV} \nonumber \\
-(3000)^2\,\text{GeV}^2 <& m^2_{H_u} &< 0 \nonumber \\
0 <& m^2_{H_d} &< (10000)^2\,\text{GeV} \nonumber \\
0<& m^2_{i_{1,2}} &< 3000 \,\text{GeV}^2 \nonumber \\
0<& m^2_{L_{3},\bar{e}_{3},\bar{d}_{3}} &< 3000 \,\text{GeV}^2 \nonumber \\
-(1000)^2 \,\text{GeV}^2 <& m^2_{Q_{3},\bar{u}_{3}} &< (1000)^2 \,\text{GeV}^2 \nonumber \\
-3000\,\text{GeV} <& A_t,A_b,A_{\tau} &< 3000\,\text{GeV} \nonumber \\
1 <& \tan \beta &< 50 \nonumber \\
&sign(\mu)&=\pm 1\, \nonumber\\
&y_t \in [1,3]&.
\end{eqnarray}
The plots in Figure \ref{QFPplots} confirm our expectation with significant reduction in fine-tuning observed in models with quasi-fixed point running of $m^2_{H_u}$ and $\mu$. In particular, low-sensitivity towards $\Lambda$ ($\Delta \lesssim \mathcal{O}(100)$) can even be maintained in models where $\Lambda$ is as high as $10^{19}$ GeV. In particular we find $\Delta^{QFP}_{min}=29$ for $\Lambda = 10^{10}$ GeV within all constraints, with similar results for $\Lambda=10^{16},10^{19}$ GeV.

\section{Conclusions} \label{concl}

The major conclusion we would like to draw in this paper is that the naturalness considerations within supersymmetric theories in light of current experimental data may be indicating towards physics beyond the MSSM that enters at scales as low as $\Lambda \sim 100$ TeV. To demonstrate this point we have treated the MSSM as an effective theory below the scale $\Lambda$ without any a priori assumption on soft-breaking parameters at $\Lambda$. The general scan of 20 MSSM parameters shows a reduction of the fine-tuning measure from $\Delta \sim \mathcal{O}(100)$ for $\Lambda \simeq 10^{16}$ GeV down to $\Delta \sim \mathcal{O}(10)$ for $\Lambda \simeq 10^{5}$ GeV, even for a sparticle spectrum lying in the multi-TeV region. We have also argued, that theories with a special quasi-fixed point behaviour of running parameters may also have reduced ($\Delta <  \mathcal{O}(100)$) fine tuning. Our results call for further exploration of non-standard theories beyond the MSSM. 

\paragraph{Acknowledgement.}
We would like to thank Lei Wu for useful discussions. The work was supported by the Australian Research Council.


\newpage

\begin{thebibliography}{999}
  
  \bibitem{gluino-atlas}
  M. Aaboud \textit{et al.}  [ATLAS Collaboration],
  Phys. Rev. D \textbf{96} 112010 (2017).
  
  \bibitem{Ross:2017kjc}
  G.~G.~Ross, K.~Schmidt-Hoberg and F.~Staub,
  JHEP {\bf 1703} (2017) 021.


  \bibitem{Barbieri1988} 
  R. Barbieri and G. Giudice, 
  Nucl. Phys. B \textbf{306} (1988) 63-76.
  
  
\bibitem{Cassel:2009ps}
  S.~Cassel, D.~M.~Ghilencea and G.~G.~Ross,
  Nucl.\ Phys.\ B {\bf 825} (2010) 203.
  
  \bibitem{casas2015}
  J. A. Casas \textit{et. al.},
  JHEP \textbf{06} (2015) 070.
  
   \bibitem{Hill1981}
C. T. Hill,
Phys. Rev. D \textbf{24} (1981) 691.

\bibitem{Ross1981}
B. Pendleton, G.G. Ross,
Phys. Lett. B. \textbf{98} (1981) 291-294.

  \bibitem{Allanach1997}
B. C. Allanach, S.A. Abel,
Phs. Lett. B, \textbf{415} (1997) 371-382.

\bibitem{Yeghi1999}
G. K. Yeghiyan \textit{et. al.}
Mod. Phys. Lett. A\textbf{14} (1999) 601-619.

\bibitem{Jurcisin1999}
M. Jurcisin, D.I. Kazakov, 
Mod. Phys. Lett. A\textbf{14} (1999) 671-688.

   \bibitem{Spheno2003}
W. Porod,
Comput. Phys. Commun. \textbf{153} (2003) 275-315.

\bibitem{Staub2014}
F. Staub,
Comput. Phys. Commun. \textbf{185} (2014) 1773-1790.

   \bibitem{Belanger2002}
G. Belanger \textit{et. al.},
Comput. Phys. Commun. \textbf{149} (2002) 103-120.

  \bibitem{pdg}
  K.~A.~Olive {\it et al.} [Particle Data Group Collaboration],
  Chin.\ Phys.\ C {\bf 38} 090001 (2014).
  
  \bibitem{higgs-atlas}
  G.~Aad {\it et al.}  [ATLAS Collaboration],
  Phys.\ Lett.\ B{\bf 716} (2012) 1-29.

\bibitem{higgs-cms}
  S.~Chatrchyan {\it et al.}  [CMS Collaboration],
  Phys.\ Lett.\ B {\bf 716} (2012) 30-61.

\bibitem{Calibbi:2014lga}
  L.~Calibbi, J.~M.~Lindert, T.~Ota and Y.~Takanishi,
  JHEP {\bf 1411} (2014) 106.
  
  \bibitem{planck}
  P.~A.~R.~Ade {\it et al.} [Planck Collaboration],
  arXiv:1502.01589 [astro-ph.CO].
  
  \bibitem{Xenon2017}
   E. Aprile \textit{et. al.} [Xenon Collaboration],
   Phys. Rev. Lett. \textbf{119}, 181301 (2017).
   
   \bibitem{Xenon2018}
   E. Aprile \textit{et. al.} [Xenon Collaboration],
   arXiv:1805.12562 [astro-ph.CO].
   
      \bibitem{Bechtle2015}
   P. Bechtle \textit{et. al.},
   Eur. Phys. Journal C \textbf{75} (2015) 421.
   
   \bibitem{HFAG2011}
   D. Asner \textit{et. al.} [Heavy Flavor Averaging Group (HFAG)],
   arXiv:1010.1589 [hep-ex].
   
   \bibitem{CMSLHCb2011}
   CMS and LHCb collaborations,
   CMS PAS BPH-11-019 (2011).
   
         \bibitem{Flav1}
   W. Porod, F. Staub and A. Vicente, 
   A. Eur. Phys. J. C, \textbf{74} (2014) 2992.
   
   \bibitem{Baker2017}
   M.J. Baker, J. Kopp,
   Phys. Rev. Lett. \textbf{119} 061801 (2017).
   
   \bibitem{Kobakhidze2017}
   A. Kobakhidze, M.A. Schmidt, M. Talia,
   arXiv:1712.05170 [hep-ph].
   
  \bibitem{Dermisek2006}
R. Dermisek, H.D. Kim, K. Ian-Woo,
JHEP \textbf{2006} (2006) 001.
  
  \bibitem{Dermi2006}
  R. Dermisek, H. D. Kim,
  Phys. Rev. Lett., \textbf{96} 211803 (2006).
  
  \bibitem{Arvan2014}
  A. Arvanitaki \textit{et. al.},
  JHEP \textbf{1403} (2014) 022.

\end{thebibliography}
\end{document}